\documentclass[a4paper]{jpconf}
\usepackage{graphicx}
\usepackage{amsmath,amssymb}
\usepackage{slashed}

\newcommand*\Bell{\text{\boldmath$\ell$\unboldmath}}

\begin{document}
\title{Angular distributions in pion-nucleon Drell-Yan process}

\author{I~V~Anikin$^1$, L~Szymanowski$^2$, O~V~Teryaev$^1$ and \underline{N~Volchanskiy}$^{1,3}$}

\address{$^1$ Bogoliubov Laboratory of Theoretical Physics, JINR, 141980 Dubna, Russia}
\address{$^2$ National Centre for Nuclear Research (NCBJ),
            00-999 Warsaw, Poland}
\address{$^3$ Research Institute of Physics, Southern Federal University,
             344090 Rostov-on-Don, Russia}

\ead{nikolay.volchanskiy@gmail.com}

\begin{abstract}
We study the Drell-Yan process in the polarized pion-nucleon scattering. We calculate the $U(1)_\text{em}$ gauge invariant hadron tensors. To this end, we have to take into account two diagrams---the standard diagram and an additional one---both diagrams are required to get the gauge invariant hadron tensor. We obtain a new single transverse spin asymmetry that probes gluon poles and chiral-odd distribution functions. The process of pion-nucleon scattering in the limit of large Feynman $x$ is described by a convolution of two non-collinear exclusive-channel distribution amplitudes, which demonstrates the duality between different factorization regimes.
\end{abstract}

\section{Introduction}

For decades the Drell-Yan (DY) process has been remaining one of the most convenient objects in the studies of hadron structure, both from the theoretical and experimental sides. Several facilities are currently carrying out or developing programs dedicated to measuring DY processes---SPS (COMPASS) \cite{Baum:1996yv, COMPASS}, RHIC \cite{Bai:2013plv}, NICA \cite{Kouznetsov:2017bip, Savin:2016arw}. An important family of observables in these experiments is single transverse spin asymmetries (SS$_\perp$A), which are related to the three-dimensional hadron structure owing to the inextricable link between the transverse spin and the parton transverse momentum dependence \cite{Angeles-Martinez:2015sea, Boer:2011fh, Boer:2003cm, Kang:2011hk, Boer:2011fx, Arnold:2008kf}.

In this report based mostly on Ref.~\cite{ASTV17}, we apply ideas developed in \cite{Anikin:2010wz, Anikin:2015xka, Anikin:2015esa} (see also talk given by I. Anikin in this workshop) to DY process induced by the polarized pion-nucleon scattering at large Feynman $x$. It turns out that the inclusion of gluon poles in the spirit of \cite{Anikin:2010wz, Anikin:2015xka, Anikin:2015esa} predicts a new SS$_\perp$A non-trivially dependent on the azimuthal angle of nucleon spin vector (in the Collins--Soper frame). Moreover, as we will see in the following, the non-trivial angular dependence of SS$_\perp$A can be considered as a signal of the parton transverse momentum dependence.

On the other hand, the limit of large Feynman $x$ allows us to describe pions by wave functions and distribution amplitudes rather than parton distributions. In a sense, this reveals a duality between semi-inclusive and exclusive processes or, put in other words, transverse momentum and generalized parton distributions. Besides, it gives a possibility to work with the gluon poles in the most explicit form---represented by a hard gluon propagator rather than inclusive function $B^V(x_1,x_2)$.

\section{Kinematics. Sudakov decompositions}

\begin{figure}[h]
\begin{minipage}{0.36\linewidth}
\includegraphics[width=\linewidth,trim=200 580 200 100,clip]{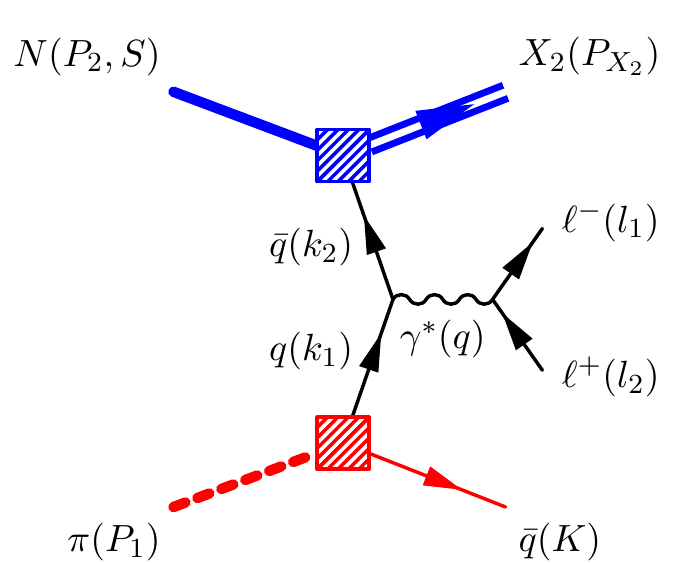}
\caption{\label{fig:dyprocess}Pion-nucleon Drell-Yan process: the lowest-order diagram. In the case we consider, the nucleon is transversely polarized, leptons are unpolarized.}
\end{minipage}\hspace{0.05\linewidth}
\begin{minipage}{0.59\linewidth}
\centering
\begin{picture}(70,140)
\put(-15,0){\includegraphics[width=0.55\linewidth]{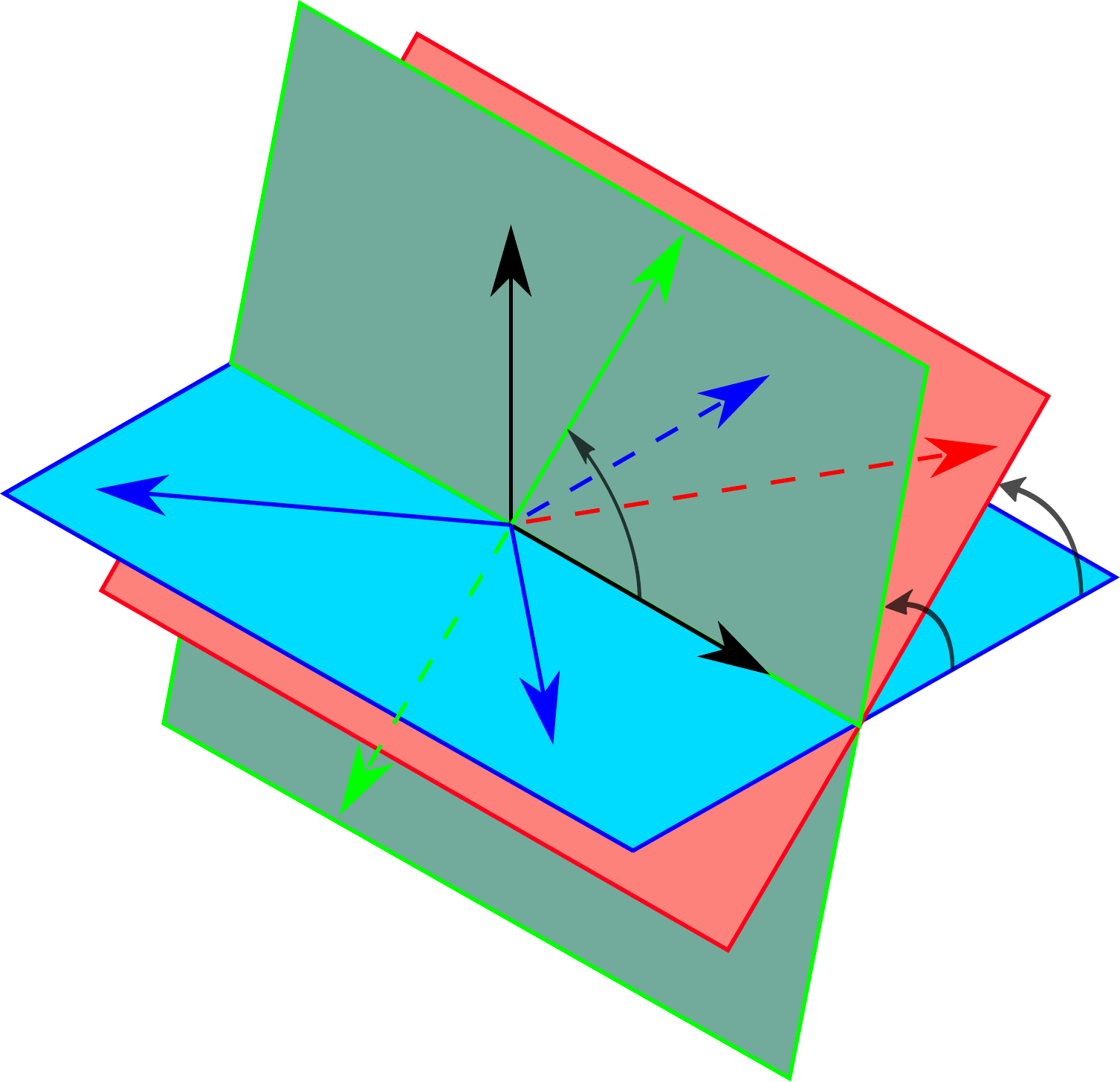}}
\put(25,25){$ \mathbf l_2$}
\put(75,115){$ \mathbf l_1$}
\put(10,80){$ \mathbf P_1$}
\put(60,37){$ \mathbf P_2$}
\put(109,86){$  \mathbf S$}
\put(80,95){$\hat{ \mathbf x }$}
\put(49,116){$\hat{ \mathbf  y}$}
\put(83,60){$\hat{ \mathbf  z}$}
\put(111,61){$\phi$}
\put(127,75){$\varphi_{S}$}
\put(70,74){$\theta$}
\end{picture}
\hspace{0.3\linewidth}
\begin{picture}(70,130)
\put(-20,0){\includegraphics[width=0.4\linewidth]{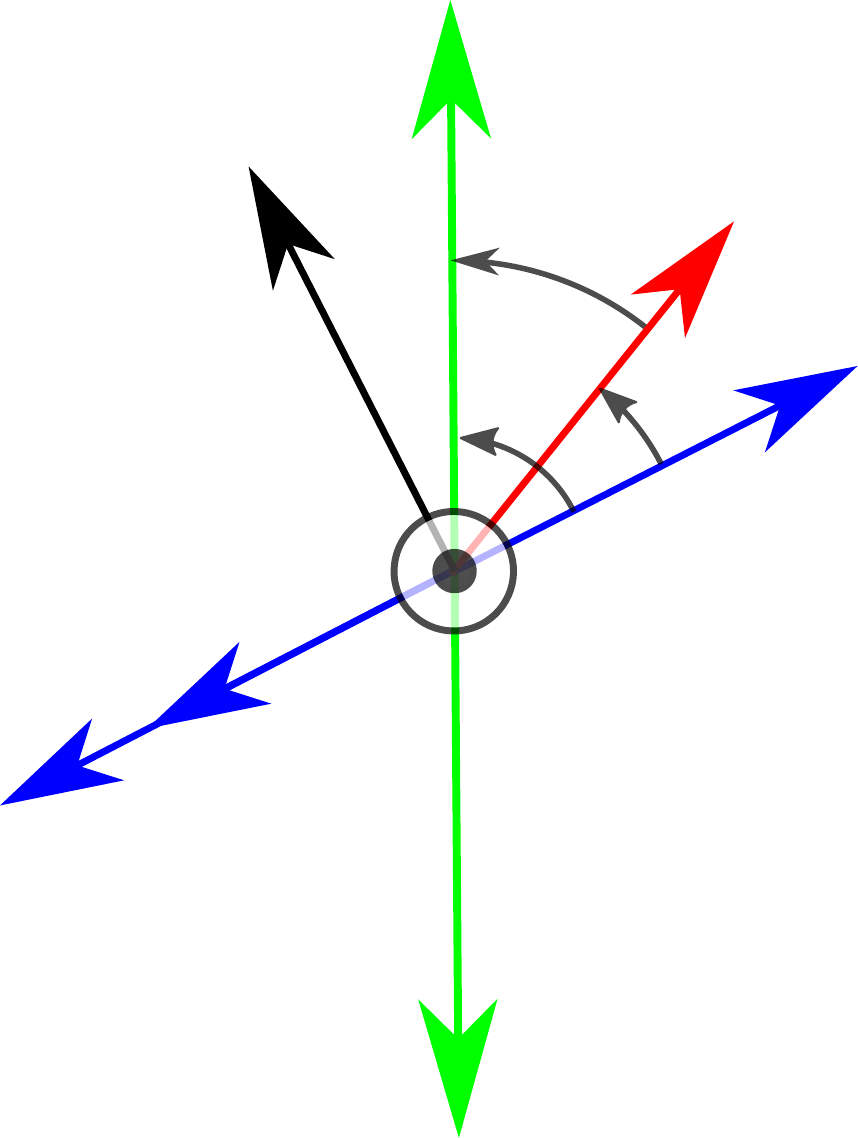}}
\put(45,5){$ \mathbf l_2$}
\put(45,130){$ \mathbf l_1$}
\put(-20,55){$ \mathbf P_1$}
\put(3,42){$ \mathbf P_2$}
\put(70,117){$  \mathbf S$}
\put(80,80){$\hat{ \mathbf x }$}
\put(5,125){$\hat{ \mathbf  y}$}
\put(40,55){$\hat{ \mathbf  z}$}
\put(45,90){$\phi$}
\put(60,92){$\varphi_{S}$}
\put(45,115){$\varphi_\text{CS}$}
\end{picture}
\caption{\label{fig:csframe}Collins--Soper frame. Cyan, olive and red planes are scattering, lepton and nucleon-spin planes, respectively. The colour of a vector corresponds to the colour of a plane the vector lies in.}
\end{minipage} 
\end{figure}

We study DY process in pion-nucleon scattering (Fig.~\ref{fig:dyprocess}) with nucleon $N$ being transversely polarized, $\pi(P_1) + N(P_2,S) \to \gamma^*(q) + \bar q(K) + \text{spectators} \to \ell(l_1)+\bar\ell(l_2) + \bar q(K) + \text{spectators}$. The virtual photon produces a pair of unpolarized leptons with an invariant mass squared $q^2=Q^2$ assumed to be large enough to make use of the factorization theorem for the process.

The Sudakov decompositions of momenta and nucleon spin 4-vector read
\begin{gather}
\label{Sudakov-decom-1}
P_1\approx \frac{Q}{x_B \sqrt{2}}\, n^*  + P_{1\,\perp} \, , \quad
P_2\approx \frac{Q}{y_B \sqrt{2}}\, n  + P_{2\,\perp} \, , \quad
S\approx \frac{\lambda}{M_N}\,P_2 + S_\perp\,
\\
\label{Sudakov-decom-2}
q=\frac{Q}{\sqrt{2}}\, (n^* + n) + q_{\perp},\, \quad q_\perp^2\ll Q^2\, ,
\end{gather}
where $\lambda$ and $M_N$ are nucleon helicity and mass, $x_B$ and $y_B$ are usual dimensionless Bjorken variables, $n^*$ and $n$ are plus and minus light-cone vectors,
\begin{gather}
n^{*\mu} = (1/\sqrt2,0,0,1/\sqrt2),
\quad
n^{\mu} = (1/\sqrt2,0,0,-1/\sqrt2).
\end{gather}

In what follows, all Lorentz non-invariant quantities are defined in the Collins--Soper frame \cite{Arnold:2008kf, Barone:2001sp}, which is a rest frame of the dilepton pair with unit vectors of the spatial axes
\begin{align}
\hat{ \mathbf x } \sim -\hat{ \mathbf P }_1 - \hat{ \mathbf P }_2,
\qquad
\hat{ \mathbf z } \sim -\hat{ \mathbf P }_1 + \hat{ \mathbf P }_2,
\qquad
\hat{ \mathbf y } = \hat{ \mathbf z } \times \hat{ \mathbf x }.
\end{align}
Here, a hatted vector  is a vector of the same direction and unit magnitude, $\hat{\mathbf{A}}=\mathbf{A}/|\mathbf{A}|$. All relevant angles are depicted in Fig.~\ref{fig:csframe}---the polar $\theta$ and azimuthal $\phi$ angles of the lepton $\ell^-(l_1)$, the angle $\varphi_S$ between scattering plane and nucleon-spin plane (azimuthal angle of the nucleon spin vector).

\section{Hadron tensor}

The cross section of the DY process can be written \cite{Arnold:2008kf, Barone:2001sp} as
\begin{eqnarray}
\label{dsigma}
\frac{\mathrm{d}\sigma}{\mathrm{d}^4 q \, \mathrm{d}\Omega_{\mathbf{k}_1}^\text{CS}}=\frac{\alpha^2_\text{em}}{2 Q^4 s}
{L}_{\mu\nu} W^{\mu\nu},
\end{eqnarray}
where $s=(P_1+P_2)^2$; $\Omega_{\mathbf{k}_1}^\text{CS}$ is the solid angle of lepton in the Collins--Soper frame; $L_{\mu\nu}$ and $W_{\mu\nu}$ stand for the lepton and hadron tensor, respectively. Since the leptons are unpolarized in the case we consider, the lepton tensor is real. As a consequence, the hadron tensor has to be real as well.

\begin{figure}[h]
\centering
\begin{picture}(400,150)
\put(0,15){\includegraphics[width=0.45\textwidth]{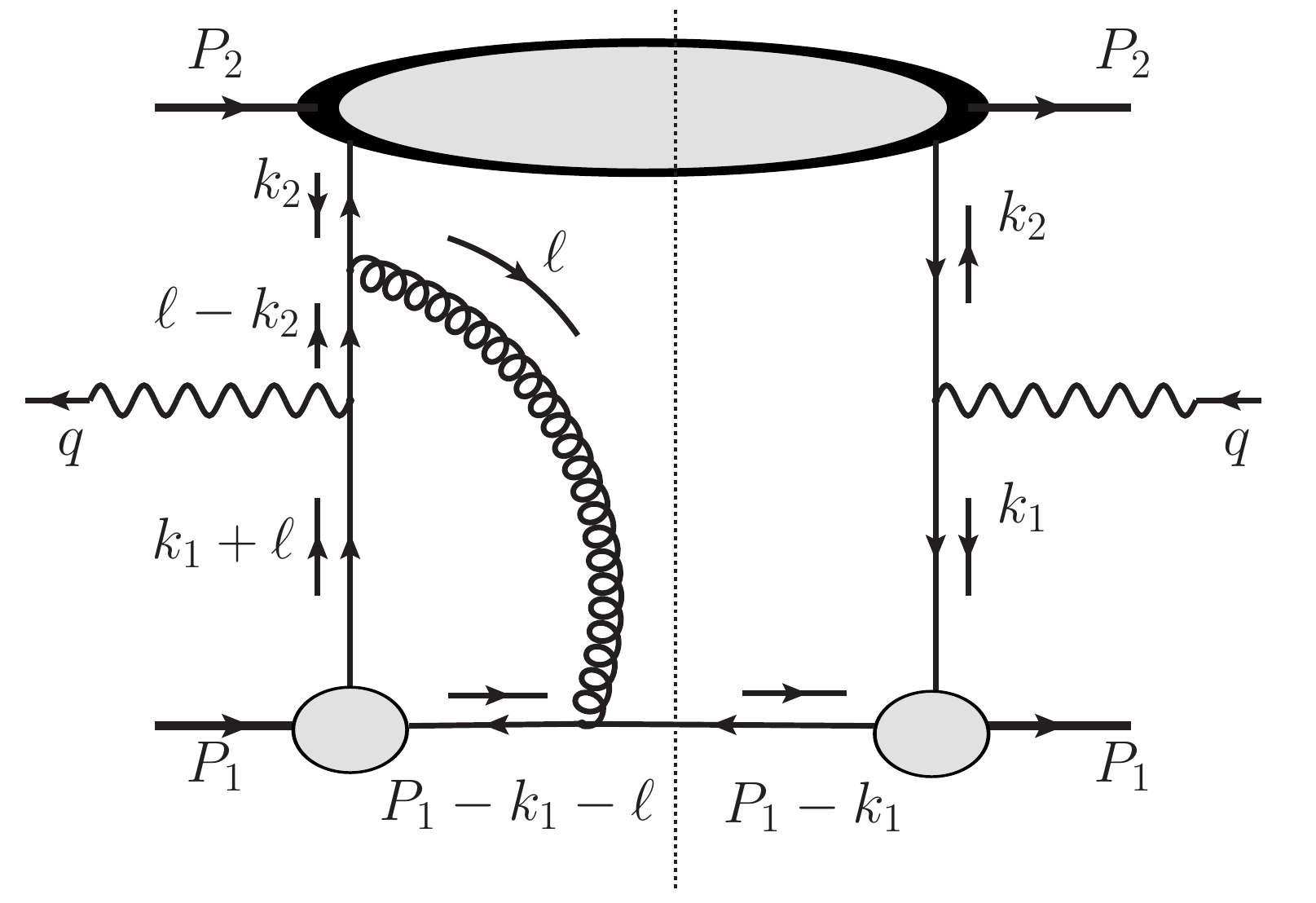}}
\put(47,5){(a) standard diagram}
\put(200,15){\includegraphics[width=0.45\textwidth]{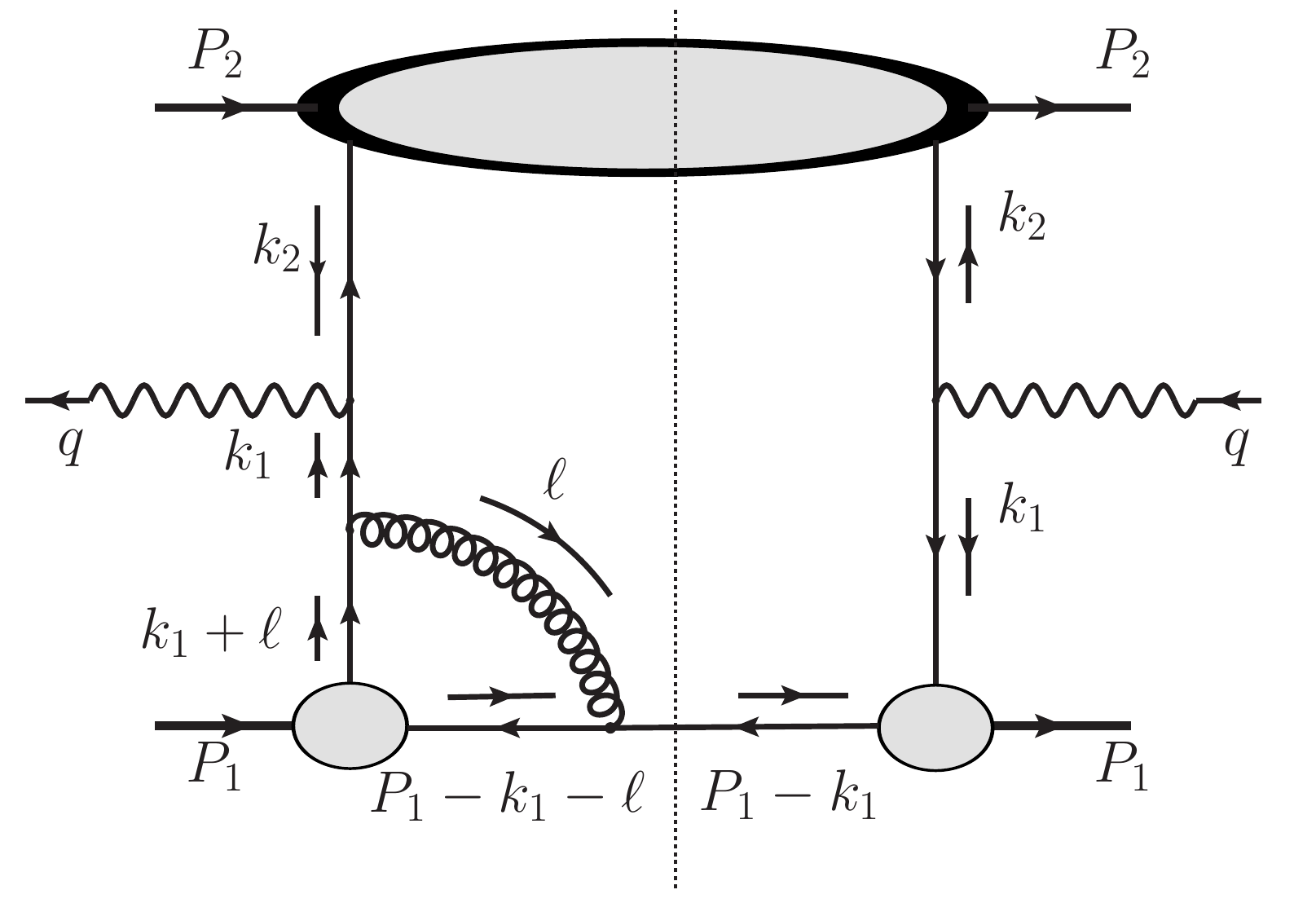}}
\put(237,5){(b) non-standard diagram}
\end{picture}
\caption{\label{Fig-DY-1-2-a}The Feynman diagrams contributing to the polarized DY hadron tensor: the standard diagram (left) and non-standard one (right).}
\end{figure}

Before the collinear factorization \cite{Anikin:2010wz, Anikin:2015xka, Anikin:2015esa}, the standard (Fig.~\ref{Fig-DY-1-2-a}a) and non-standard (Fig.~\ref{Fig-DY-1-2-a}b) diagrams give the following contributions to the hadron tensor:
\begin{multline}
\label{HadTen-St}
W_{\mu\nu}^{(\text{stand.})} \sim \int \mathrm d^4k_1\,\mathrm d^4k_2 \, \delta^{(4)}(k_1+k_2-q)
\\
\times\int \mathrm d^4\ell \, D_{\alpha\beta}(\ell)
\Tr \big[ \gamma_\nu \Gamma \gamma_\alpha S(\ell-k_2) \gamma_\mu \Gamma_1 \gamma_\beta \Gamma_2
\big]
\bar\Phi^{[\Gamma]}(k_2)\,
\Phi_{(2)}^{[\Gamma_1]}(k_1; \ell)\,
\Phi_{(1)}^{[\Gamma_2]}(k_1)\, \delta\big( (P_1-k_1)^2 \big) \,,
\end{multline}
\begin{multline}
\label{HadTen-NonSt}
W_{\mu\nu}^{(\text{nonstand.})} \sim \int \mathrm d^4k_1\,\mathrm d^4k_2 \, \delta^{(4)}(k_1+k_2-q)
\\
\times\int \mathrm d^4\ell\, D_{\alpha\beta}(\ell)
\Tr\big[ \gamma_\nu \Gamma \gamma_\mu S(k_1) \gamma_\alpha \Gamma_1 \gamma_\beta \Gamma_2
\big]\bar\Phi^{[\Gamma]}(k_2)\,
\Phi_{(2)}^{[\Gamma_1]}(k_1; \ell)\,
\Phi_{(1)}^{[\Gamma_2]}(k_1)\, \delta\big( (P_1-k_1)^2 \big),
\end{multline}
where
\begin{align}\label{barPhi-func}
\bar\Phi^{[\Gamma]}(k_2)&{}= \int\frac{\mathrm d^4\eta_2}{(2\pi)^4} e^{-ik_2\eta_2}\,
\langle P_2, S|\bar\psi(\eta_2) \Gamma \psi(0)\big] | S , P_2\rangle\, ,
\\ \label{Phi2-func}
\Phi_{(2)}^{[\Gamma_1]}(k_1; \ell)&{}=
\int \frac{\mathrm d^4\eta_1}{(2\pi)^4} e^{i(P_1-\ell-k_1)\eta_1}\, \langle 0 | \bar\psi(\eta_1) \,\Gamma_1 \psi(0) | P_1\rangle\,,
\\ \label{Phi1-func}
\Phi_{(1)}^{[\Gamma_2]}(k_1)&{}= \int \frac{\mathrm d^4\xi}{(2\pi)^4} e^{-ik_1\xi}\,
\langle P_1 | \bar\psi(\xi) \,\Gamma_2 \psi(0) | 0\rangle\,.
\end{align}
In  Eqns.~(\ref{HadTen-St}) and (\ref{HadTen-NonSt}), the $\delta$-functions $\delta\big( (P_1-k_1)^2 \big)$ indicate that the quark with the momentum $P_1-k_1$ is on its mass shell.

Let us  consider first the standard contribution, which exists even for real $B^V$-function. We constrain ourselves to the case with one of the pion distribution amplitudes being projected onto the matrix related to the chiral-odd distribution. Then, the pion chiral-odd distribution amplitudes ``automatically'' isolates the chiral-odd structure in the nucleon matrix element. So, the matrices $\Gamma$, $\Gamma_1$, $\Gamma_2$ and related distribution amplitudes in Eqn.~(\ref{HadTen-St}) are as follows:
\begin{eqnarray}\label{gamma-str}
\begin{gathered}
\Gamma \bar\Phi^{[\Gamma]} = \sigma^{+\perp}
 \bar\Phi^{[\sigma^{-\perp} ]},
\qquad
\Gamma_1 \Phi_{(2)}^{[\Gamma_1]} =  \gamma^-(\gamma_5)  \Phi_{(2)}^{[\gamma^+(\gamma_5)]},
\qquad
\Gamma_2 \Phi_{(1)}^{[\Gamma_2]} =
 \sigma^{+ -} (\gamma_5) \Phi_{(1)}^{[\sigma^{- +}(\gamma_5) ]}.
\end{gathered}
\end{eqnarray}
These are the twist-2 nucleon parton distribution and the pion distribution amplitudes of twist two and three. Following the factorization procedure (see, for example, \cite{An-ImF, Belitsky:2005qn, Diehl:2003ny, Braun:2011dg}), we have for the upper blob in Fig.~\ref{Fig-DY-1-2-a}a
\begin{align}
\label{Upper-Phi}
\bar\Phi^{[\sigma^{-\perp}]}(y)\stackrel{\text{def.}}{=}
\int(\mathrm d^4k_2) \delta(y-k_2^-/P_2^-) \bar\Phi^{[\sigma^{-\perp}]}(k_2) = \varepsilon^{-\perp S^{\perp} P_2}
\bar h_{1}(y).
\end{align}
For the lower blobs in Fig.~\ref{Fig-DY-1-2-a}a, we obtain a following convolution
\begin{align}
\label{Lower-Phi-2}
\mathbb{B}_{\alpha\beta}^{[\Gamma_2,\,\Gamma_1]}(x_1,x_2)=
\frac{g^\perp_{\alpha\beta}}{2P_1^+ (x_2-x_1)}
\Phi_{(1)}^{[\Gamma_2]}(x_1)\hspace{-0.2cm}
\stackrel{\hspace{0.2cm}{\bf k}_{1}^{\,\perp}}{\circledast}
\hspace{-0.1cm}\Phi_{(2)}^{[\Gamma_1]}(\bar x_{2})\,,
\end{align}
where the integration over $\mathrm dk^-_1$ has  been done with the help of $\delta((P_1-k_1)^2)$ and the convolution over the transverse momentum is defined as
\begin{gather}
\label{Conv-Delta}
F \stackrel{\hspace{0.2cm}{\bf k}_{1}^{\,\perp}}{\circledast} G =
\int(\mathrm d^2{\bf k}_{1}^{\,\perp})  \, F({\bf k}_{1}^{\,\perp})
\int(\mathrm d\ell^- \mathrm d^2\Bell_\perp) \, \Delta(\ell^-,{\Bell}_\perp)\,
G({\bf k}_{1}^\perp;\ell^- \Bell_\perp)\,,
\\
\Delta(\ell^-,\Bell_\perp)= \frac{1}{\ell^- - {\Bell}^{\,2}_\perp/(2(x_2-x_1)P_1^+) + i\,{\rm sign}(x_2-x_1)\, 0}.
\end{gather}

The convolution of the two non-collinear pion distribution amplitudes, the function $\mathbb{B}_{\alpha\beta}^{[\Gamma_2,\,\Gamma_1]}(x_1,x_2)$ of Eqn.~(\ref{Lower-Phi-2}), is an analogue of the function $B^V(x_1,x_2)$ that is introduced for the description of the inclusive channel \cite{Anikin:2010wz}. The representation through two non-collinear functions is physically sound, if the invariant mass of pion-induced spectators is not large. In the limit of $\lvert \Bell_\perp\rvert\gg \lvert{\bf k}_{1}^{\,\perp}\rvert$, we can write a similar representation through two collinear distribution amplitudes. The resemblance between the functions $B^V(x_1,x_2)$ and $\mathbb{B}_{\alpha\beta}^{[\Gamma_2,\,\Gamma_1]}(x_1,x_2)$ can be seen as a manifestation of duality between exclusive and inclusive channels.

In Eqn.~(\ref{Lower-Phi-2}), the gluon pole at $x_1=x_2$ can be conveniently dealt with in the contour gauge (see Refs.~\cite{Anikin:2010wz, Anikin:2015xka, Anikin:2015esa} and the talk given by I.~Anikin in this workshop). This leads to a modification
\begin{eqnarray}\label{GP}
\frac{1}{x_2-x_1} \stackrel{\text{c. g.}}{\Longrightarrow} \frac{1}{x_2-x_1 - i\epsilon},
\end{eqnarray}
which follows from the integral representation of the unit step function \cite{Anikin:2015xka}.

The non-standard term in the hadron tensor (Fig.~\ref{Fig-DY-1-2-a}b) can be considered along the same lines as the standard one \cite{Anikin:2010wz}. Completing all proper decompositions and simplifications, we finally get the following expressions for the hadron tensor:
\begin{multline}
\label{HadTen-St-2}
 W_{\mu\nu}^{(\text{stand.})} \sim i \int \mathrm dx_1\,\mathrm dy \, \delta^{(4)}(x_1P+yP_2-q)
\overline{\Phi}^{[\Gamma]}(y)
\\
\times\int \mathrm dx_2\,
\Tr\big[ \gamma_\nu \Gamma \gamma_\alpha \frac{(x_2-x_1)\slashed P_1^+}{-(x_2-x_1) y s + i0}
\gamma_\mu \gamma^- \gamma_\beta \Gamma_2
\big]
\frac{1}{2}\,\frac{g^\perp_{\alpha\beta} \tilde n^{[\,-} P_1^{+\,]}}{x_2-x_1 - i\epsilon}
\Phi_{(1)}^{\text{tw-3}}(x_1)\hspace{-0.2cm} \stackrel{\hspace{0.2cm}{\bf k}_{1}^{\,\perp}}{\circledast}
\hspace{-0.1cm}\Phi_{(2)}^{\text{tw-2}}(x_2)\,,
\end{multline}
\begin{multline}
\label{HadTen-NonSt-2}
W_{\mu\nu}^{(\text{nonstand.})} \sim i\int \mathrm dx_1\,\mathrm dy \, \delta^{(4)}(x_1P_1+yP_2-q)
\overline{\Phi}^{[\Gamma]}(y)
\\
\times
\Tr\big[ \gamma_\nu \Gamma \gamma_\mu \frac{\gamma^+}{2x_1P_1^++i0} \gamma_\alpha \gamma^- \gamma_\beta \Gamma_2
\big] \frac{1}{2}\int \mathrm dx_2\frac{g^\perp_{\alpha\beta}\, \tilde n^{[\,-} P_1^{+\,]}}{x_2-x_1 - i\epsilon}
\Phi_{(1)}^{\text{tw-3}}(x_1)\hspace{-0.2cm} \stackrel{\hspace{0.2cm}{\bf k}_{1}^{\,\perp}}{\circledast}
\hspace{-0.1cm}\Phi_{(2)}^{\text{tw-2}}(x_2)\,.
\end{multline}

It can be readily seen that both terms~(\ref{HadTen-St-2}) and (\ref{HadTen-NonSt-2}) are not gauge invariant in QED sense, while the sum thereof is invariant. The sum reads
\begin{multline}
\label{HadTen-GI}
\overline{W}^{\mu\nu}= \int \mathrm d^2 {\bf q}_\perp { W}^{\mu\nu}
\\
\sim i \int \mathrm dx_1\, \mathrm dy \, \delta(x_1P^+_1-q^+)\delta(yP^-_2-q^-)
\bar h_{1}(y) \int \mathrm dx_2 \widetilde{B}(x_1,x_2) \,\frac{\varepsilon^{\nu S_\perp P_2 P_1}_\perp}{(P_1 P_2)}
\Big[ \frac{P^{\mu}_1}{y}- \frac{P^{\mu}_2}{x_1}\Big] \,,
\end{multline}
where
\begin{eqnarray}
\label{B-fun}
\widetilde{B}(x_1,x_2)=\frac{1}{2}
\frac{\Phi_{(1)}^{\text{tw-3}}(x_1)\hspace{-0.2cm}
\stackrel{\hspace{0.2cm}{\bf k}_{1}^{\,\perp}}{\circledast}
\hspace{-0.1cm}\Phi_{(2)}^{\text{tw-2}}(x_2)}{x_2-x_1 - i\epsilon}\,.
\end{eqnarray}
The above expression coincides formally with the result obtained in \cite{Anikin:2010wz} for the inclusive Drell-Yan process.

\section{Single transverse spin asymmetry}

SS$_\perp$A is written as
\begin{eqnarray}
\label{SSA-1}
{\cal A}= \frac{ \mathrm{d}\sigma^{(\uparrow)} -  \mathrm{d}\sigma^{(\downarrow)} }{  \mathrm{d}\sigma^{(\uparrow)} +  \mathrm{d}\sigma^{(\downarrow)} },
\end{eqnarray}
where $\mathrm{d}\sigma^{(\uparrow)}$ and $\mathrm{d}\sigma^{(\downarrow)}$ are cross sections \eqref{dsigma} with opposite directions of the nucleon transverse spin. For the process under consideration, we can obtain the SS$_\perp$A from Eqns.~\eqref{HadTen-GI} and \eqref{B-fun}. This asymmetry for the chiral-odd contribution can be put in terms of \cite{COMPASS} as follows:
\begin{eqnarray}
\label{NewSSA}
{\cal A}=\frac{|\mathbf S_{\perp}|}{Q} \, \frac{  D_{[\sin 2\theta]} \, \sin\phi_S B^{\sin\phi_S}_{UT}}
{\bar f_1(y_B)\,H_1(x_B)},
\qquad
 D_{[\sin 2\theta]}=\frac{\sin 2\theta}{1+\cos^2\theta},
\end{eqnarray}
where $B^{\sin\phi_S}_{UT}=2\bar h_1(y_B)\, \Phi_{(1)}^{\text{tw-3}}(x_B)\hspace{-0.2cm} \stackrel{\hspace{0.2cm}{\bf k}_{1}^{\,\perp}}{\circledast} \hspace{-0.1cm}\Phi_{(2)}^{\text{tw-2}}(x_B)$; $\bar f_1(y_B)$ and $H_1(x_B)$ parametrize the following matrix elements contributing to the unpolarized cross section
\begin{gather}
\label{Unpol}
\langle P_2| \bar\psi\, \gamma^- \,\psi |P_2 \rangle\stackrel{{\cal F}}{\sim} P^-_2\bar f_1(y),
\qquad
\langle P_1| \bar\psi_+ |q(K)\rangle \langle q(K)| \psi_+ |P_1 \rangle\stackrel{{\cal F}}{\sim} P_1^+ H_1(x),
\\
\label{Unpol2}
H_1(x)= 
\frac{1}{2\bar x_1 P_1^+} \int \mathrm (d^2{\bf k}_{1}^{\perp})  
\Phi^{[\gamma^+(\gamma_5)]}_{(2)}(\bar x_1,{\bf k}_{1}^{\perp}) 
\Phi^{[\mathbb{I}(\gamma_5)]}_{(1)}(x_1,{\bf k}_{1}^{\perp}).
\end{gather}

It should be stressed that the asymmetry \eqref{NewSSA} is different from both the leading twist Sivers asymmetry $A^{\sin\phi_S}_{UT}$, which is accompanied by the depolarization factor $D_{[1+\cos^2\theta]}$, and the higher twist asymmetries $A^{\sin(\phi_S\pm\phi)}_{UT}$, which are related to a different tensor structure, although appearing with the same factor $D_{[\sin 2\theta]}$ as the asymmetry \eqref{NewSSA}.

\section{Conclusion}

We derived the hadron tensor for pion-nucleon DY process with large Feynman $x$, transversely polarized nucleon and ``primordial'' transverse momenta. We took into account two diagrams (Fig.~\ref{Fig-DY-1-2-a}), standard diagram and non-standard \cite{Anikin:2010wz} one---both diagrams are necessary to get manifestly $U(1)_\text{em}$ gauge invariant hadron tensor. The pion-related soft part of the diagrams is described by a convolution of two non-collinear distribution functions---a sort of decomposition of the inclusive function $B^V(x_1,x_2)$. This approximation implies that the invariant masses of undetected spectators in the pion sector are relatively small. We studied the case of the nucleon distribution and one of the pion amplitudes being chiral-odd. In this case, we obtained unexpected angular dependence of SS$_\perp$A that can be measured experimentally. The asymmetry can be seen as a signal for the transverse ``primordial'' momentum dependence probing both gluon poles and chiral-odd functions and related to the duality of different factorization regimes in exclusive and inclusive channels (see, \cite{Anikin:2008bq}).

\ack

We thank A.V.~Efremov, L.~Motyka and A.~P.~Nagaytsev for useful discussions. The work by I.V.A. was partially supported by the Bogoliubov--Infeld Program and Heisenberg--Landau Program HL-2017. L.Sz. is partly supported by grant No $2015/17/$B/ST$2/01838$ from the National Science Center in Poland.


\section*{References}

\begin{thebibliography}{99}

\bibitem{Baum:1996yv}
  Baum G {\it et al} [COMPASS Collaboration] 1996
  COMPASS: A proposal for a common muon and proton apparatus for structure and spectroscopy
  CERN-SPSLC-96-14 CERN-SPSLC-P-297

\bibitem{COMPASS}
  Aghasyan M {\it et al} [COMPASS Collaboration] 2017
  {First measurement of transverse-spin-dependent azimuthal asymmetries in the Drell-Yan process} 
  {\it Preprint} arXiv:1704.00488 [hep-ex].

\bibitem{Bai:2013plv}
  Bai M {\it et al} 2013 Status and plans for the polarized hadron collider at RHIC {\it Proc. 4th Int. Particle Accelerator
                        Conference (IPAC 2013): Shanghai, China, May 12-17, 2013} TUYB103

\bibitem{Kouznetsov:2017bip}
  Kouznetsov O and Savin I 2017
  \textit{Nucl.\ Part.\ Phys.\ Proc.}  {\bf 282} 20

\bibitem{Savin:2016arw}
  Savin I {\it et al} 2016
  {\it Eur.\ Phys.\ J.} A {\bf 52} 215

\bibitem{Angeles-Martinez:2015sea}
  Angeles-Martinez~R {\it et al} 2015
  {\it Acta Phys.\ Polon.} B {\bf 46} 2501

\bibitem{Boer:2011fh}
  Boer D {\it et al} 2011
  {Gluons and the quark sea at high energies: Distributions, polarization, tomography}
  \textit{Preprint} arXiv:1108.1713 [nucl-th].

\bibitem{Boer:2003cm}
  Boer D, Mulders P J and Pijlman F 2003
  \textit{Nucl.\ Phys.} B {\bf 667} 201

\bibitem{Kang:2011hk}
  Kang Z B, Qiu J W, Vogelsang W and Yuan F 2011
  \textit{Phys.\ Rev.} D {\bf 83} 094001

\bibitem{Boer:2011fx}
  Boer D 2011
  \textit{Phys.\ Lett.} B {\bf 702} 242

\bibitem{Arnold:2008kf}
  Arnold S, Metz A and Schlegel M 2009
  {\it Phys.\ Rev.} D {\bf 79} 034005

\bibitem{ASTV17}
  Anikin~I~V, Szymanowski~L and Teryaev~O~V and Volchanskiy~N 2017
  {\it Phys.\ Rev.} D {\bf 95} 111501


\bibitem{Anikin:2010wz}
  Anikin I V and Teryaev O V 2010
  {\it Phys.\ Lett.} B {\bf 690} 519 

\bibitem{Anikin:2015xka}
  Anikin I V and Teryaev O V 2015
  {\it Eur.\ Phys.\ J.} C {\bf 75} 184 

\bibitem{Anikin:2015esa}
  Anikin I V and Teryaev O V 2015
  {\it Phys.\ Lett.} B {\bf 751} 495 

\bibitem{Barone:2001sp}
  Barone V, Drago A and Ratcliffe P G 2002
  {\it Phys.\ Rept.}  {\bf 359} 1 

\bibitem{An-ImF}
  Anikin I V, Ivanov~D~Y, Pire~P, Szymanowski~L and Wallon~S 2010
  {\it Nucl.\ Phys.}  B {\bf 828} 1 

\bibitem{Belitsky:2005qn}
  Belitsky~A~V and Radyushkin~A~V 2005
  {\it Phys.\ Rept.}  {\bf 418} 1 

\bibitem{Diehl:2003ny}
  Diehl~M 2003
  {\it Phys.\ Rept.}  {\bf 388} 41 

\bibitem{Braun:2011dg}
  Braun~V~M and Manashov~A~N 2012
  {\it J. High Energy Phys.} JHEP01(2012)085 

\bibitem{Anikin:2008bq}
  Anikin I V, Cherednikov~I~O, Stefanis~N~G and Teryaev O V 2009
  {\it Eur.\ Phys.\ J.} C {\bf 61} 357 

\end{thebibliography}
\end{document}